\documentclass[aps, prc, floatfix, nofootinbib, superscriptaddress, twocolumn]{revtex4-1}

\usepackage{latexsym}
\usepackage{amsmath}
\usepackage{amssymb}
\usepackage{amsfonts}
\usepackage{makecell}

\usepackage[mathscr,scaled=1.15]{urwchancal}
\DeclareFontFamily{OT1}{pzc}{}
\DeclareFontShape{OT1}{pzc}{m}{it}%
{<-> s * [1.15] pzcmi7t}{}
\DeclareMathAlphabet{\mathpzc}{OT1}{pzc}{m}{it}

\usepackage{color}

\usepackage{supertabular}
\usepackage{placeins}
\usepackage{epsfig}
\usepackage{graphicx}
\usepackage{multirow}

\definecolor{purple}{rgb}{0.5,0,0.5}
\definecolor{blue}{rgb}{0.0,0,0.9}
\definecolor{prdblue}{rgb}{0.133,0.118,0.498}
\usepackage[colorlinks=true, pdfstartview=FitV, linkcolor=prdblue, citecolor= prdblue, urlcolor=prdblue]{hyperref}

\begin{document}

\title{Study on $Z_{cs}$ and excited $B_s^0$ states in the chiral quark model}

\author{Xiaoyun Chen}
\email[]{xychen@jit.edu.cn}
\affiliation{College of Science, Jinling Institute of Technology, Nanjing 211169, P. R. China}

\author{Yue Tan}
\email[]{181001003@njnu.edu.cn}
\affiliation{Department of Physics, Nanjing Normal University, Nanjing 210023, P. R. China}

\author{Yuan Chen}
\email[]{cy1208@nuaa.edu.cn}
\affiliation{Key Laboratory of Radio Frequency and Micro-Nano Electronics of Jiangsu Province, Nanjing 210023, P. R. China}

\begin{abstract}
Stimulated by the newly observed charged hidden-charm state $Z_{cs}(3985)^-$ by BESIII Collaboration, $Z_{cs}(4000)^+$, $Z_{cs}(4220)^+$
and the excited $B_s^0$ states by LHCb Collaboration, a full calculation including masses and decay widths is emerged in the chiral quark model. For $Z_{cs}$ states, we assign quantum numbers $I(J^P)=\frac{1}{2}(1^+)$ and quark composition $c\bar{c}s\bar{u}$ according to the experiment. For $B_s^0$ states, systematically investigations are performed with $I(J^P)=0(0^+), 0(1^+), 0(2^+)$ both in 2-body $b\bar{s}$ and  4-body $b\bar{s}q\bar{q}~(q=u~\rm{or}~d)$ systems. Each tetraquark calculation takes all structures including meson-meson, diquark-antidiquark and all possible color configurations into account. Among the numerical techniques to solve the 2-body and 4-body Schr{\"o}dinger equation, the spatial wave functions are expanded in series of Gaussian basis functions for high precision, which is the way Gaussian expansion method so called. Our results indicate that the low-lying states of 4-quark system are all higher than the corresponding thresholds either for $c\bar{c}s\bar{u}$ or for $b\bar{s}q\bar{q}$ systems. With the help of the real scaling method, we found two resonance states with masses of 4023 MeV and 4042 MeV for $c\bar{c}s\bar{u}$ system. The state $c\bar{c}s\bar{u}(4042)$ has a consistent mass and decay width with the recent observed state $Z_{cs}(3985)^-$. For $b\bar{s}q\bar{q}$ system with $J=0$, some resonance states are also found. The newly observed excited $B_s^0$ states can be accommodated in the chiral quark model as $2S$ or $1D$ states, and the mixing with four-quark states are also needed to be considered.
\end{abstract}

\maketitle

%%%%%%%%%%%%%%%%%%%%%%%%%%%%%%%%%%%%%%%%%%%%%%%%%%%%%%%%%%%%%%%%%%%%%%%%%%%%%%%%%%%%%%%%%%%%%%%%%%%%%%%%%%%%%%%%%%%%%%%

\section{Introduction} \label{introduction}
Recently, for the first time, the BESIII Collaboration has reported a structure $Z_{cs}(3985)^-$ in the $K^+$ recoil-mass spectrum
near the $D_s^-D^{*0}$ and $D_s^{*-}D^0$ mass thresholds in the process of $e^+e^- \rightarrow K^+(D_s^-D^{*0}+D_s^{*-}D^0)$ at the
center-of-mass energy $\sqrt{s}=4.681$ GeV, with the mass and narrow decay width~\cite{Ablikim:2020hsk},
\begin{eqnarray}
   M_{Z_{cs}}&=& (3982_{-2.6}^{+1.8}\pm2.1)\quad\rm{MeV}, \nonumber\\
   \Gamma_{Z_{cs}}&=&(12.8_{-4.4}^{+5.3}\pm3.0)\quad\rm{MeV},
\end{eqnarray}
and the significance was estimated to be 5.3$\sigma$. Soon afterwards, a significant state $Z_{cs}(4000)^+$, with a mass of
$4003\pm6^{+4}_{-14}$ MeV, a width of $131\pm15\pm26$ MeV, and spin-parity $J^P = 1^+$, was observed by LHCb Collaboration,
including another exotic state $Z_{cs}(4220)^+$ \cite{Aaij:2021ivw}. The discovery of the charged heavy quarkonium-like structures
with strangeness could shed light on the properties of the charged exotic $Z$ states reported before~\cite{Zcfamily}.

Besides $Z_{cs}$ states, two excited $B_s^0$ states are observed in the $B^+K^-$ mass spectrum in a sample of proton-proton collisions at
centre-of-mass energies of 7, 8, and 13 TeV very recently by LHCb Collaboration~\cite{Aaij:2020hcw}. The masses and widths of the two states
are determined to be
\begin{eqnarray}
   M_1&=& 6063.5\pm1.2(\rm{stat})\pm0.8(\rm{syst})~\rm{MeV}, \nonumber\\
   \Gamma_1&=&26\pm4(\rm{stat})\pm4(\rm{syst})~\rm{MeV},  \\
    M_2&=& 6114.5\pm3(\rm{stat})\pm5(\rm{syst})~\rm{MeV}, \nonumber\\
   \Gamma_2&=&66\pm18(\rm{stat})\pm21(\rm{syst})~\rm{MeV}.
\end{eqnarray}

For $Z_{cs}$, it is classified into the exotic state as the strange partner of $Z_c(3900)$ and has intensively attracted more attentions and
investigations theoretically within a very short time
\cite{Meng:2020ihj,Wang:2020htx,Azizi:2020zyq,Sungu:2020zvk,Wang:2020rcx,Chen:2020yvq,Yang:2020nrt,Jin:2020yjn,Ferretti:2020ewe,Wang:2020kej}.
These explanations basically cover various exotic hadron configurations. One feature of the $Z_{cs}$ is that its mass is on the verge of the
$\bar{D}_sD^*$ or $\bar{D}_s^*D$ threshold, so a molecular resonance is suggested. For example, Lu Meng $\emph{et al.}$ obtained the mass and
width of $Z_{cs}$ in good agreement with the experimental results by considering the coupled-channel effect and strongly supported the $Z_{cs}$
states as the $U/V$-spin partner states of the charged $Z_c(3900)$~\cite{Meng:2020ihj}. In chiral effective field theory up to the
next-to-leading order, $Z_{cs}$ also was regarded as the partner of the $Z_c(3900)$ in the SU(3) flavor symmetry and the
$\bar{D}_sD^*/D_s^*D$ molecular resonance~\cite{Wang:2020htx}. In the QCD sum rule, $Z_{cs}$ can be well defined as a diquark-antidiquark
candidate with quark content $\bar{c}cu\bar{s}$~\cite{Sungu:2020zvk}.

On the contrary, J. Ferretti pointed out that the meson-meson molecular model could not be used to describe heavy-light tetraquarks with
non-null strangeness content, and in the case of $cs\bar{c}\bar{n}~(n = u~{\rm or}~d$) configurations, the compact tetraquark ground-state
is about 200 MeV below the lowest energy hadro-charmonium state, $\eta_cK$~\cite{Ferretti:2020ewe}. Another explanation is that $Z_{cs}$
can be naturally regarded as a reflection structure from a charmed-strange meson $D_{s2}^*(2573)$ by Lanzhou group~\cite{Wang:2020kej}.
By adopting a one-boson-exchange model and considering the coupled channel effect, Ref.~\cite{Chen:2020yvq} excluded $Z_{cs}$ as a
$D^{*0}D_s^-/D^0D_s^{*-}/D^{*0}D_s^{*-}$ resonance. As more and more exotic states named as $XYZ$ have been observed in different experiments,
their structures are still inexplicable and controversial theoretically. Investigating for charged charmonium-like states can extend our
knowledge of hadrons and our understanding of the nature of strong interaction. So we believe that the study of the newly observed $Z_{cs}$
states in the chiral quark model can provide some useful information on exotic hadrons. In present work, for $Z_{cs}$, the quantum number is
assigned as $I(J^P)=\frac{1}{2}(1^+)$ and the quark composition is $c\bar{c}s\bar{u}$.

Now let's turn to the newly observed excited $B_s^0$ states. In the past few years, many experiment collaborations such as CDF, D0, and LHCb
have made contributions to find the radial and orbital excitations of the bottom and bottom-strange meson families. More and more higher
excitations emerged in experiments~\cite{Aaltonen:2013atp,Aaltonen:2007ah,Abazov:2007af,Aaij:2012uva}. Therewith many theoretical studies of
the bottom and bottom-strange mesons follow close on another~\cite{Sun:2014wea,Xiao:2014ura,Liu:2015lka,Godfrey:2016nwn,Lu:2016bbk,Ferretti:2015rsa}.
The mass spectrum and strong decay patterns are studied most in conventional 2-body quark-antiquark system, which can describe the ground states
very well, but has poor understanding for higher excitations of bottom and bottom-strange mesons. Studying the $B$ and $B_s$ mesons will help us
not only understand of excited mesons, but also put the discovered excited charm and chram-strange mesons into the larger context, since
the present situation of experimental exploration of bottom and bottom-strange states is very similar to that of $D$ and $D_s$ states in 2003
\cite{Aubert:2003fg,Besson:2003cp,Krokovny:2003zq,Evdokimov:2004iy,Aubert:2006mh,Aubert:2009ah,delAmoSanchez:2010vq}.
With newly observation of the excited $B_s^0$ states by LHCb Collaboration~\cite{Aaij:2020hcw}, now it is a good time to carry out a
comprehensive theoretical study on higher bottom-strange mesons. In this work, all possible quantum numbers with $I(J^P)=0(0^+), 0(1^+),
0(2^+)$ are studied for $B_s^0$ states. Considering the possible limitation of quark-antidiquark system in conventional quark model
in describing the higher excitations and the possible production of quark-antiquark pair in the vacuum, we obtain the masses of $B_s^0$ states
in 2-body quark-antidiquark system and 4-body $b\bar{s}q\bar{q}~(q=u~\rm{or}~d)$ system, respectively. $b\bar{s}s\bar{s}$ is not included here
because of its high energy.

Each tetraquark calculation takes into account the mixing of structures, such as meson-meson and diquark-antidiquark structure, along with
all possible color, spin configurations. In the meantime, in order to find possible stable resonance states, high precision computing method
Gaussian expansion method (GEM) \cite{GEM} and an useful stabilization real scaling method are both employed~\cite{rs1,rs2} in our calculations.

The paper is arranged as follows. Theoretical framework including the chiral quark model, the wave functions of $Z_{cs}$ and $B_s^0$,
along with GEM are introduced in Section \ref{ModleandGEM}. In Section \ref{results}, the numerical results and discussion are presented.
A short summary is given in Section \ref{epilogue}.
%%%%%%%%%%%%%%%%%%%%%%%%%%%%%%%%%%%%%%%%%%%%%%%%%%%%%%%%%%%%%%%%%%%%%%%%%%%%%%%%%%%%%%%%%%%%%%%%%%%%%%%%%%%%%%%%%%%%%%%

\section{Theoretical framework}
\label{ModleandGEM}
\emph{\textbf{Chiral quark model:}} the review of the chiral quark model and GEM has been introduced in
Refs.~\cite{Chen:2018hts,Chen:2017mug,Chen:2016npt}, and here they will be introduced briefly and we mainly focus on the relevant features of
$Z_{cs}$ and $B_s^0$ states.

The Hamiltonian of the chiral quark model can be written as follows for 4-body system,
\begin{align}
 H & = \sum_{i=1}^4 m_i  +\frac{p_{12}^2}{2\mu_{12}}+\frac{p_{34}^2}{2\mu_{34}}
  +\frac{p_{1234}^2}{2\mu_{1234}}  \quad  \nonumber \\
  & + \sum_{i<j=1}^4 \left[ V_{ij}^{C}+V_{ij}^{G}+\sum_{\chi=\pi,K,\eta} V_{ij}^{\chi}
   +V_{ij}^{\sigma}\right].
\end{align}
The potential energy: $V_{ij}^{C, G, \chi, \sigma}$ represents the confinement, one-gluon-exchange, Goldston boson exchange and $\sigma$
exchange, respectively. The detailed forms can be referred to Eq. (13) in Ref.~\cite{Chen:2017mug}, which are omit here for space
saving. All the model parameters are determined by fitting the meson spectrum, from light to heavy; and the resulting values are listed in
 Table~\ref{modelparameters}.

\begin{table}[!t]
\begin{center}
\caption{ \label{modelparameters}Model parameters, determined by fitting the meson spectrum.}
\begin{tabular}{llr}
\hline\noalign{\smallskip}
Quark masses   &$m_u=m_d$    &313  \\
   (MeV)       &$m_s$         &536  \\
               &$m_c$         &1728 \\
               &$m_b$         &5112 \\
\hline
Goldstone bosons   &$m_{\pi}$     &0.70  \\
   (fm$^{-1} \sim 200\,$MeV )     &$m_{\sigma}$     &3.42  \\
                   &$m_{\eta}$     &2.77  \\
                   &$m_{K}$     &2.51  \\
                   &$\Lambda_{\pi}=\Lambda_{\sigma}$     &4.2  \\
                   &$\Lambda_{\eta}=\Lambda_{K}$     &5.2  \\
                   \cline{2-3}
                   &$g_{ch}^2/(4\pi)$                &0.54  \\
                   &$\theta_p(^\circ)$                &-15 \\
\hline
Confinement        &$a_c$ (MeV fm$^{-2}$)         &101 \\
                   &$\Delta$ (MeV)     &-78.3 \\
\hline
OGE                 & $\alpha_0$        &3.67 \\
                   &$\Lambda_0({\rm fm}^{-1})$ &0.033 \\
                  &$\mu_0$(MeV)    &36.98 \\
                   &$s_0$(MeV)    &28.17 \\
\hline
\end{tabular}
\end{center}
\end{table}

It is to be noted that, only $V^{\chi=\eta}$ of Goldston boson exchange plays a role between $u$ and $s$ quark, and for $u$ and $\bar{u}$
interacting quark-pair, not only $V^{\chi=\pi,\eta}$, but also $V^{\sigma}$ works; for other quark pairs such as, $(Q,u)$, $(Q,s)$, $(Q,Q)$
($Q=c,b$), only $V_{ij}^{C, G}$ is considered without Goldstone bosons and $\sigma$ exchange.

\begin{figure}
\resizebox{0.50\textwidth}{!}{\includegraphics{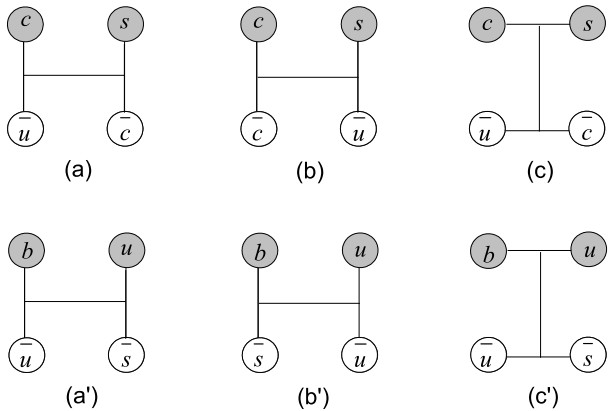}}
\caption{\label{structure} Structures of $Z_{cs}$ and $b\bar{s}q\bar{q}$ system, $(a)/(a^{\prime})$ and $(b)/(b^{\prime})$ represent
two meson-meson structures, and $(c)/(c^{\prime})$ represents diquark-antidiquark structure.}
\end{figure}

\emph{\textbf{Wave functions:}} There are three quark configurations for $Z_{cs}$ and $b\bar{s}q\bar{q}$ system , two meson-meosn structures and one diquark-antidiquark structure, which are shown in Fig.\ref{structure}.
For spin part, the wave functions for 2-body system are
\begin{align}
&\chi_{11}=\alpha\alpha,~~
\chi_{10}=\frac{1}{\sqrt{2}}(\alpha\beta+\beta\alpha),~~
\chi_{1-1}=\beta\beta,\nonumber \\
&\chi_{00}=\frac{1}{\sqrt{2}}(\alpha\beta-\beta\alpha),
\end{align}
then total six wave functions of 4-body system are obtained easily, which are shown in Table \ref{wavefunctions}(the first column). The subscripts $SM_S$ of $\chi$ represents the total spin and the third projection of total spin of four-quark system, with $S=0,1,2$, and only one component ($M_S=S$) is shown for a given total spin $S$.
For flavor part, the flavor wave functions (two meson-meson structures plus one diquark-antidiquark structure) are also tabulated in Table \ref{wavefunctions}(the second column). The wave functions $\chi^{f1}$, $\chi^{f2}$, $\chi^{f3}$ correspond to the picture $(a),(b),(c)$ in Fig \ref{structure} for $Z_{cs}$, respectively, and the $\chi^{f4}$, $\chi^{f5}$, $\chi^{f6}$ are the wave functions of pictures $(a^{\prime}),(b^{\prime}),(c^{\prime})$ for $b\bar{s}q\bar{q}$ system.
For color part, there are four wave functions in total(the third column in Table \ref{wavefunctions}), $\chi^{c1}$ color singlet-singlet $(1\otimes1)$ and $\chi^{c2}$
color octet-octet $(8\otimes8)$ for meson-meson structure and $\chi^{c3}$ color antitriplet-triplet $(\bar{3}\otimes3)$ and $\chi^{c4}$
sextet-antisextet $(6\otimes\bar{6})$ for diquark-antidiquark structure.

\linespread{1.5}
\begin{table*}[!t]
\begin{center}
\caption{ \label{wavefunctions}The wave functions of spin, flavor, color part for $Z_{cs}$ and $b\bar{s}q\bar{q}$ system by considering all
kinds of quark structures.}
\begin{tabular}{lll}
\hline\hline\noalign{\smallskip}
\quad \quad \quad Spin                                              & Flavor                   &\quad\quad \quad \quad \quad \quad \quad Color \\
\hline
$\chi_{00}^{\sigma1}=\chi_{00}\chi_{00}$
&$\chi^{f1}=c\bar{u}s\bar{c}$
&\quad \quad \quad$\chi^{c1}=\frac{1}{3}(\bar{r}r+\bar{g}g+\bar{b}b)(\bar{r}r+\bar{g}g+\bar{b}b)$    \\

$\chi_{00}^{\sigma2}=\sqrt{\frac{1}{3}}(\chi_{11}\chi_{1-1}-\chi_{10}\chi_{10}+\chi_{1-1}\chi_{11})$
&$\chi^{f2}=c\bar{c}s\bar{u}$
&\quad \quad \quad\makecell[l]{$\chi^{c2}=\frac{\sqrt{2}}{12}(3\bar{b}r\bar{r}b+3\bar{g}r\bar{r}g+3\bar{b}g\bar{g}b+3\bar{g}b\bar{b}g+3\bar{r}g\bar{g}r$\\
$\quad \quad \quad +3\bar{r}b\bar{b}r+2\bar{r}r\bar{r}r+2\bar{g}g\bar{g}g+2\bar{b}b\bar{b}b-\bar{r}r\bar{g}g$ \\
$\quad \quad \quad -\bar{g}g\bar{r}r-\bar{b}b\bar{g}g-\bar{b}b\bar{r}r-\bar{g}g\bar{b}b-\bar{r}r\bar{b}b)$}\\

$\chi_{11}^{\sigma3}=\chi_{00}\chi_{11}$
&$\chi^{f3}=cs\bar{u}\bar{c}$
&\quad \quad \quad\makecell[l]{$\chi^{c3}=\frac{\sqrt{3}}{6}(rg\bar{r}\bar{g}-rg\bar{g}\bar{r}+gr\bar{g}\bar{r}-gr\bar{r}\bar{g}$ \\
$\quad \quad \quad+rb\bar{r}\bar{b}-rb\bar{b}\bar{r}+br\bar{b}\bar{r}-br\bar{r}\bar{b}$ \\
$\quad \quad \quad+gb\bar{g}\bar{b}-gb\bar{b}\bar{g}+bg\bar{b}\bar{g}-bg\bar{g}\bar{b})$}  \\

$\chi_{11}^{\sigma4}=\chi_{11}\chi_{00}$
&$\chi^{f4}=\frac{1}{2}(b\bar{d}d\bar{s}+b\bar{u}u\bar{s})$
&\quad \quad \quad\makecell[l]{$\chi^{c4}=\frac{\sqrt{6}}{12}(2rr\bar{r}\bar{r}+2gg\bar{g}\bar{g}+2bb\bar{b}\bar{b}+rg\bar{r}\bar{g}+rg\bar{g}\bar{r}$\\
$\quad \quad \quad+gr\bar{g}\bar{r}+gr\bar{r}\bar{g}+rb\bar{r}\bar{b}+rb\bar{b}\bar{r}+br\bar{b}\bar{r}$  \\
$\quad \quad \quad+br\bar{r}\bar{b}+gb\bar{g}\bar{b}+gb\bar{b}\bar{g}+bg\bar{b}\bar{g}+bg\bar{g}\bar{b})$}\\

$\chi_{11}^{\sigma5}=\frac{1}{\sqrt{2}}(\chi_{11}\chi_{10}-\chi_{10}\chi_{11})$
&$\chi^{f5}=-\frac{1}{2}(b\bar{s}u\bar{u}+b\bar{s}d\bar{d})$
&\makecell[l]{\quad\\\quad} \\

$\chi_{22}^{\sigma6}=\chi_{11}\chi_{11}$
&$\chi^{f6}=-\frac{1}{2}(bu\bar{u}\bar{s}+bd\bar{d}\bar{s})$
& \\
\hline\hline
\end{tabular}
\end{center}
\end{table*}

\linespread{1.5}
\begin{table*}[!t]
\begin{center}
\renewcommand\tabcolsep{6.0pt} % column width adjustment
\caption{ \label{channels} Allowed channels for $Z_{cs}$ and $b\bar{s}q\bar{q}$ system, for saving context space, we give abbreviations for channels, \emph{e.g.}, for meson-meson structure (picture $(a)$) of $Z_{cs}$, $'3 1 1'$ and $'3 1 2'$ represents $\chi_{11}^{\sigma3}\chi^{f1}\chi^{c1}$ and $\chi_{11}^{\sigma3}\chi^{f1}\chi^{c2}$, severally. And $'4 1 1'$ and $'4 1 2'$ represents $\chi_{11}^{\sigma4}\chi^{f1}\chi^{c1}$ and $\chi_{11}^{\sigma4}\chi^{f1}\chi^{c2}$, respectively. The rest channels can be read in the same manner. The last row gives the total numbers of channels by considering meson-meson structures, diquark-antidiquark structure, along with all kinds of color spin configurations for $Z_{cs}$ and $b\bar{s}q\bar{q}$ system with quantum numbers $0(0^+)$, $0(1^+)$, $0(2^+)$.}
\begin{tabular}{ccccccccccccc}
\hline\hline\noalign{\smallskip}
system                    &\multicolumn{3}{c}{$Z_{cs}$}                      &\multicolumn{9}{c}{$b\bar{s}q\bar{q}$}  \\
\hline
$I(J^P)$
&\multicolumn{3}{c}{$\frac{1}{2}(1^+)$} &\multicolumn{3}{c}{$0(0^+)$} &\multicolumn{3}{c}{$0(1^+)$} &\multicolumn{3}{c}{$0(2^+)$} \\
Structure
&$(a)$ &$(b)$ &$(c)$ &$(a^{'})$ &$(b^{'})$ &$(c^{'})$ &$(a^{'})$ &$(b^{'})$ &$(c^{'})$ &$(a^{'})$ &$(b^{'})$ &$(c^{'})$ \\
\hline
\makecell[c]{channel \\ (spin $\cdot$ flavor $\cdot$ color)}
&\makecell[c]{3 1 1\\3 1 2\\4 1 1\\4 1 2\\5 1 1\\5 1 2}
&\makecell[c]{3 2 1\\3 2 2\\4 2 1\\4 2 2\\5 2 1\\5 2 2}
&\makecell[c]{3 3 3\\3 3 4\\4 3 3\\4 3 4\\5 3 3\\5 3 4}

&\makecell[c]{1 4 1\\1 4 2\\2 4 1\\2 4 2}
&\makecell[c]{1 5 1\\1 5 2\\2 5 1\\2 5 2}
&\makecell[c]{1 6 3\\1 6 4\\2 6 3\\2 6 4}

&\makecell[c]{3 4 1\\3 4 2\\4 4 1\\4 4 2\\5 4 1\\5 4 2}
&\makecell[c]{3 5 1\\3 5 2\\4 5 1\\4 5 2\\5 5 1\\5 5 2}
&\makecell[c]{3 6 3\\3 6 4\\4 6 3\\4 6 4\\5 6 3\\5 6 4}

&\makecell[c]{6 4 1\\ 6 4 2}
&\makecell[c]{6 5 1\\ 6 5 2}
&\makecell[c]{6 6 3\\ 6 6 4} \\
\hline
number of channels&6 &6 &6 &4 &4 &4 &6 &6 &6 &2 &2 &2\\
&\multicolumn{3}{c}{total 18}
&\multicolumn{3}{c}{total 12}
&\multicolumn{3}{c}{total 18}
&\multicolumn{3}{c}{total 6} \\
\hline\hline
\end{tabular}
\end{center}
\end{table*}

So we can get all allowed spin $\otimes$ flavor $\otimes$ color channels of $Z_{cs}$ and $b\bar{s}q\bar{q}$ system by taking meson-meson structures, diquark-antidiquark structure, along with all kinds of color spin configurations into account, which are shown in Table \ref{channels}.

Next, let's discuss the orbital wave functions for 4-body system. They can be obtained by coupling the orbital wave function for each
relative motion of the system,
\begin{equation}\label{spatialwavefunctions}
\Psi_{L}^{M_{L}}=\left[[\Psi_{l_1}({\bf r}_{12})\Psi_{l_2}({\bf
r}_{34})]_{l_{12}}\Psi_{L_r}({\bf r}_{1234}) \right]_{L}^{M_{L}},
\end{equation}
where $l_1$ and $l_2$ is the angular momentum of two sub-clusters, respectively. $\Psi_{L_r}(\mathbf{r}_{1234})$ is the wave function of
the relative motion between two sub-clusters with orbital angular momentum $L_r$. $L$ is the total orbital angular momentum of four-quark states.
Because of the positive parity ($P=(-1)^{l_1+l_2+L_r}$=+) for $Z_{cs}$ and $b\bar{s}q\bar{q}$, it is natural to assume that all the orbital angular
momenta are zeros. With the help of Gaussian expansion method (GEM), the spatial wave functions are expanded in series of Gaussian basis functions.
\begin{subequations}
\label{radialpart}
\begin{align}
\Psi_{l}^{m}(\mathbf{r}) & = \sum_{n=1}^{n_{\rm max}} c_{n}\psi^G_{nlm}(\mathbf{r}),\\
\psi^G_{nlm}(\mathbf{r}) & = N_{nl}r^{l}
e^{-\nu_{n}r^2}Y_{lm}(\hat{\mathbf{r}}),
\end{align}
\end{subequations}
where $N_{nl}$ are normalization constants,
\begin{align}
N_{nl}=\left[\frac{2^{l+2}(2\nu_{n})^{l+\frac{3}{2}}}{\sqrt{\pi}(2l+1)}
\right]^\frac{1}{2}.
\end{align}
$c_n$ are the variational parameters, which are determined dynamically. The Gaussian size
parameters are chosen according to the following geometric progression
\begin{equation}\label{gaussiansize}
\nu_{n}=\frac{1}{r^2_n}, \quad r_n=r_1a^{n-1}, \quad
a=\left(\frac{r_{n_{\rm max}}}{r_1}\right)^{\frac{1}{n_{\rm
max}-1}}.
\end{equation}
This procedure enables optimization of the expansion using just a small numbers of Gaussians. Finally, the complete channel wave function
$\Psi^{\,M_IM_J}_{IJ}$ for four-quark system is obtained by coupling the orbital and spin, flavor, color wave functions get in
Table \ref{channels}. At last, the eigenvalues of four-quark system are obtained by solving the Schr\"{o}dinger equation
\begin{equation}
    H \, \Psi^{\,M_IM_J}_{IJ}=E^{IJ} \Psi^{\,M_IM_J}_{IJ}.
\end{equation}
To obtain stable results in our work, the Gaussian width and Gaussian number of each inner cluster takes, $r_1=0.1~\rm{fm}$,
$r_n=2~ \rm{fm}$, $n=12$. For the relative motion between two sub-clusters $r_1=0.1~\rm{fm}$, $r_n=6~\rm{fm}$, $n=7$.

%%%%%%%%%%%%%%%%%%%%%%%%%%%%%%%%%%%%%%%%%%%%%%%%%%%%%%%%%%%%%%%%%%%%%%%%%%%%%%%%%%%%%%%%%%%%%%%%%%%%%%%%%%%%%%%%%%%%%%%

\section{Calculations and analysis}
\label{results}
In the present work, we calculated the mass spectrum of newly observed $Z_{cs}$ and excited $B_s^0$ states in the chiral quark model.
For the excited $B_s^0$ states, we firstly treat them as ordinary quark-antiquark states. Using the model parameters given in
Table \ref{modelparameters}, the convergent results of $b\bar{s}$ mass spectrum up to the second $D$-wave states in the chiral
quark model are obtained and shown in Table \ref{bs_results}, where the experimental data are also listed for comparison. Until now,
there has been only very limited experimental values on the low-lying bottom-strange mesons, which are called $B_s^0(5366)$,
$B_s^*(5415)$, $B_{s1}(5830)^0$, $B_{s2}^*(5840)^0$ \cite{PDG}. In Table \ref{bs_results},
we can see that the $2S$ and $1D$ states have masses between 6000 MeV and 6200 MeV, so it is possible that the newly observed excited $B_s^0$ are $2S$ or $1D$ states of $b\bar{s}$. However, for the excitation energy as high as 700 MeV, the excitation of the light quark-antiquark pair from vaccuum is highly favored. So considering the excited $B_s^0$ states as four-quark $b\bar{s}q\bar{q}~(q = u~\rm{or}~d )$ system is also necessary. In the following the four-quark $b\bar{s}q\bar{q}~(q = u~\rm{or}~d )$ system with quantum numbers $I(J^P)=0(0^+), 0(1^+), 0(2^+)$ is investigated.

%\linespread{1.5}
\begin{table}[!t]
\begin{center}
\renewcommand\tabcolsep{6.0pt} % 调整表格列间的宽度
\caption{ \label{bs_results} The mass spectrum of $b\bar{s}$ meson families in the chiral quark model in comparison with reference \cite{Sun:2014wea} and experimental data \cite{PDG} (unit: MeV).}
\begin{tabular}{ccccc}
\hline\hline\noalign{\smallskip}
$n^{2S+1}L_J$  &This work &Ref. \cite{Sun:2014wea}  &Expt \cite{PDG}\\
\hline
$1^1S_0$  &5367.4  &5390  &5366.84$\pm$0.15\\
$1^3S_1$  &5410.2  &5447  &5415.8$\pm$1.5\\
$2^1S_0$  &6017.3  &5985  &\\
$2^3S_1$  &6057.2  &6013  &\\
$1^3P_0$  &5749.2  &5830  &\\
$1^3P_1$  &5779.3  &5859  &5828.65$\pm$0.24 \\
$1^3P_2$  &5812.0  &5875  &5839.92$\pm$0.14 \\
$1^1P_1$  &5797.6  &5858  &\\
$2^3P_0$  &6345.9  &6279  &\\
$2^3P_1$  &6381.9  &6291  &\\
$2^3P_2$  &6422.9  &6295  &\\
$2^1P_1$  &6403.9  &6284  &\\
$1^3D_1$  &6179.3  &6181  &\\
$1^3D_2$  &6145.3  &6185  &\\
$1^3D_3$  &6094.2  &6178  &\\
$1^1D_2$  &6128.2  &6180  &\\
$2^3D_1$  &6778.1  &6542  &\\
$2^3D_2$  &6743.9  &6542  &\\
$2^3D_3$  &6692.9  &6534  &\\
$2^1D_2$  &6726.8  &6536  &\\
\hline\hline
\end{tabular}
\end{center}
\end{table}

For $Z_{cs}$, the minimal quark component should be $c\bar{c}s\bar{u}$ rather than a pure $c\bar{c}$ since it is observed as a charged particle
with strangeness. Both of $Z_{cs}$ and $b\bar{s}q\bar{q}$ states have two kinds of meson-meson structures and one diquark-antidiquark structure,
which are shown in Fig. \ref{structure}. Along with all possible color and spin configurations, we take all kinds of structures into account.
Table \ref{mesonmass} gives the masses of some relevant quark-antiquark mesons in present work in the chiral quark model.
From the table, we can see that the chiral quark model is very successful in describing the meson spectra. And the mass spectra of
$Z_{cs}$ and $b\bar{s}q\bar{q}$ system are demonstrated in Table \ref{4q-results}.

%\linespread{1.5}
\begin{table}[!t]
\begin{center}
\caption{ \label{mesonmass}The masses of some relevant mesons in present work in the chiral quark model, compared with the experimental data (unit: MeV).}
\begin{tabular}{cccccccc}
\hline\hline\noalign{\smallskip}
Meson  &$D^0$  &$D^{*0}$ &$D_s^*$ &$D_s^-$ &$\eta_c$ &$J/\psi$ &$K^{\pm}$ \\
  \hline
$M_{cal}$ &1862.4&1980.5   &2079.9  &1952.6  &3102.2   &3161.2  &493.9 \\
$M_{exp}$ &1864.8&2006.9   &2112.2  &1968.3  &2983.6   &3096.9  &493.6\\
\hline
Meson  &$B^-$  &$B^{*-}$ &$\bar{B_s}^0$ &$\bar{B_s}^*$ &$\omega$ &$\eta$  &$K^{*\pm}$\\
  \hline
$M_{cal}$ &5280.9 &5319.6 &5367.9 &5410.2 &701.5 &669.2  &913.6\\
$M_{exp}$ &5279.3 &5325.2 &5366.7 &5415.4 &782.6 &547.8  &891.6\\
\hline\hline
\end{tabular}
\end{center}
\end{table}

%\linespread{1.5}
\begin{table*}[!t]
\begin{center}
\renewcommand\tabcolsep{5.5pt} % 调整表格列间的宽度
\caption{ \label{4q-results}The mass spectrum of $Z_{cs}$ and $b\bar{s}q\bar{q}$ system with allowed quantum numbers. $E(a)$, $E(b)$, $E(c)$ represents the energies in pure meson-meson structures and pure diquark-antidiquark structure for $Z_{cs}$, corresponding to figure $(a)$, $(b)$, and $(c)$ in Fig. \ref{structure}. $E(a)\otimes E(c)$, $E(b)\otimes E(c)$, $E(a)\otimes E(b)$ are the energies considering the mixture of one meson-meson structure and one diquark-antidiquark structure, or two meson-meson structures, severally. $E_{cc}$ represents ground state energy for each state after considering the coupling of all
possible quark structures, color and spin channels (refer to Table \ref{channels}). It is the same with $b\bar{s}q\bar{q}$. The last column gives the theoretical lowest thresholds (unit: MeV).}
\begin{tabular}{cccccccccc}
\hline\hline\noalign{\smallskip}
& &$E(a)$   &$E(b)$   &$E(c)$ &$E(a)\otimes E(c)$ &$E(b)\otimes E(c)$ &$E(a)\otimes E(b)$ &$E_{cc}$   &the lowest thresholds \\
\hline
$Z_{cs}$    &$\frac{1}{2}(1^+)$
  &3934.5   &3656.3   &4247.1 &3934.5            &3655.4            &3934.5            &3656.2 &$3655.1(J/\psi K^-)$\\
& &$E(a^{\prime})$ &$E(b^{\prime})$ &$E(c^{\prime})$ &$E(a^{\prime})\otimes E(c^{\prime})$ &$E(b^{\prime})\otimes E(c^{\prime})$   &$E(a^{\prime})\otimes E(b^{\prime})$ &$E_{cc}$   &the lowest thresholds \\
$b\bar{s}q\bar{q}$  &$0(0^+)$  &5776.6       &6040.2       &6283.2   &5776.5&6040.2&5776.6    &5775.3     &$5774.8(B^-K^+)$\\
                    &$0(1^+)$  &5815.4       &6072.6       &6316.9   &5815.3&6072.6&5815.4    &5814.1           &$5813.5(B^{*-}K^+)$ \\
                    &$0(2^+)$  &6234.8       &6114.9       &6483.7   &6234.7&6114.9&6234.8    &6114.3           &$6111.7(\bar{B_s}^*\omega)$\\
\hline\hline
\end{tabular}
\end{center}
\end{table*}

\begin{figure}
\center{\includegraphics[width=7.0cm]{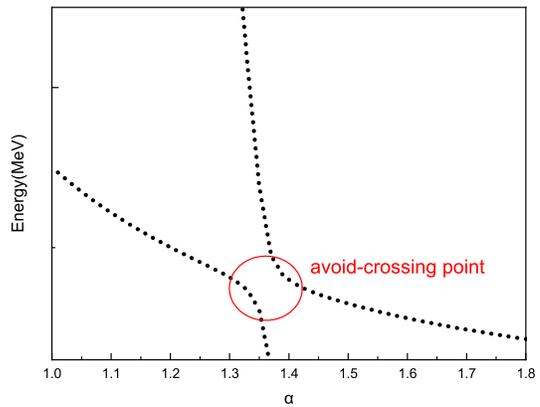}}
\caption{\label{avoid-crossing} Stabilization graph for the resonance.}
\end{figure}

\begin{figure}
\resizebox{0.50\textwidth}{!}{\includegraphics{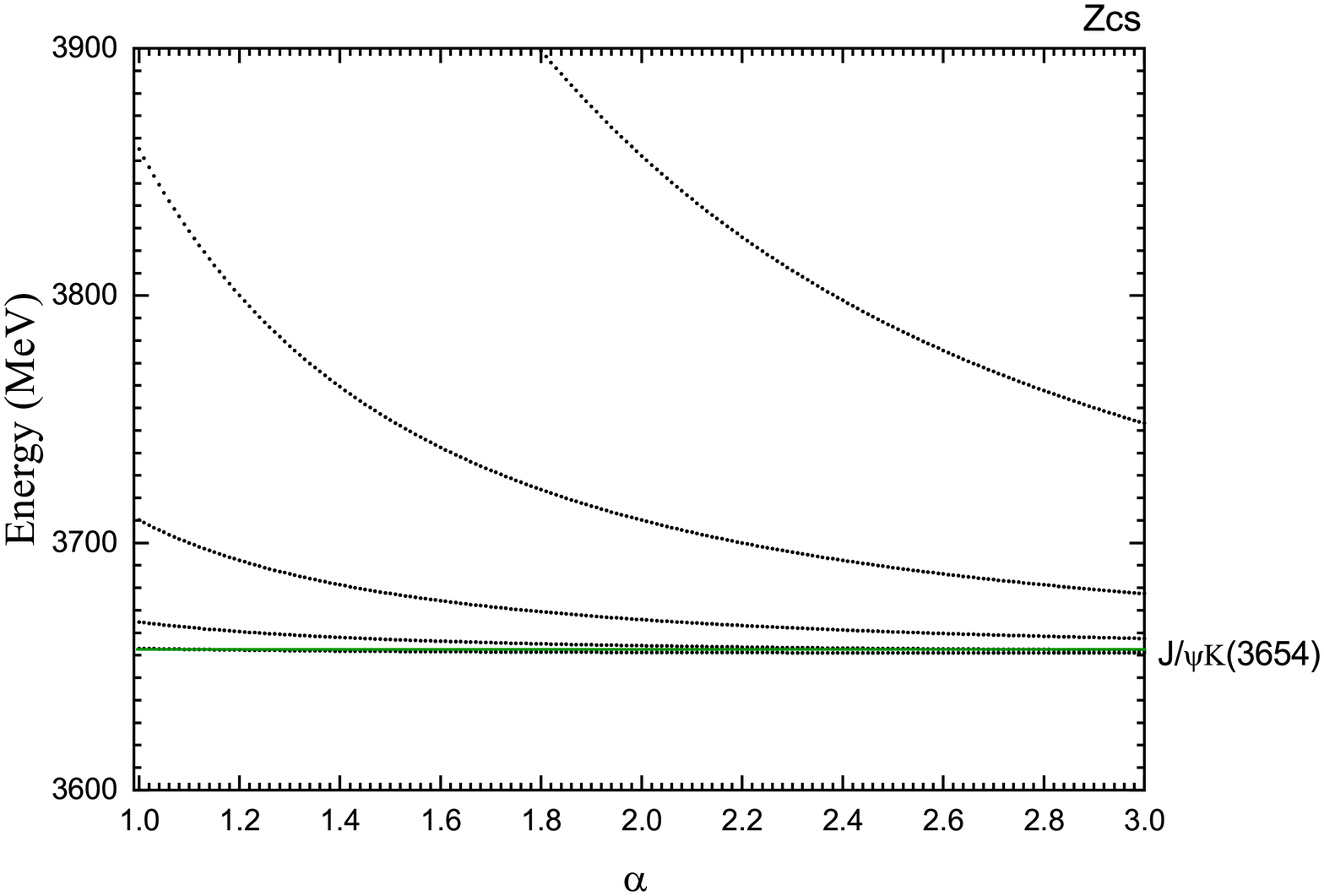}}
\caption{\label{Zcs1} The stabilization plots of the energies (3600 MeV $\sim$ 3900 MeV) of $c\bar{c}s\bar{u}$ states for $I(J^P)=\frac{1}{2}(1^+)$ with respect to the scaling factor $\alpha$.}
\end{figure}

\begin{figure}
\resizebox{0.50\textwidth}{!}{\includegraphics{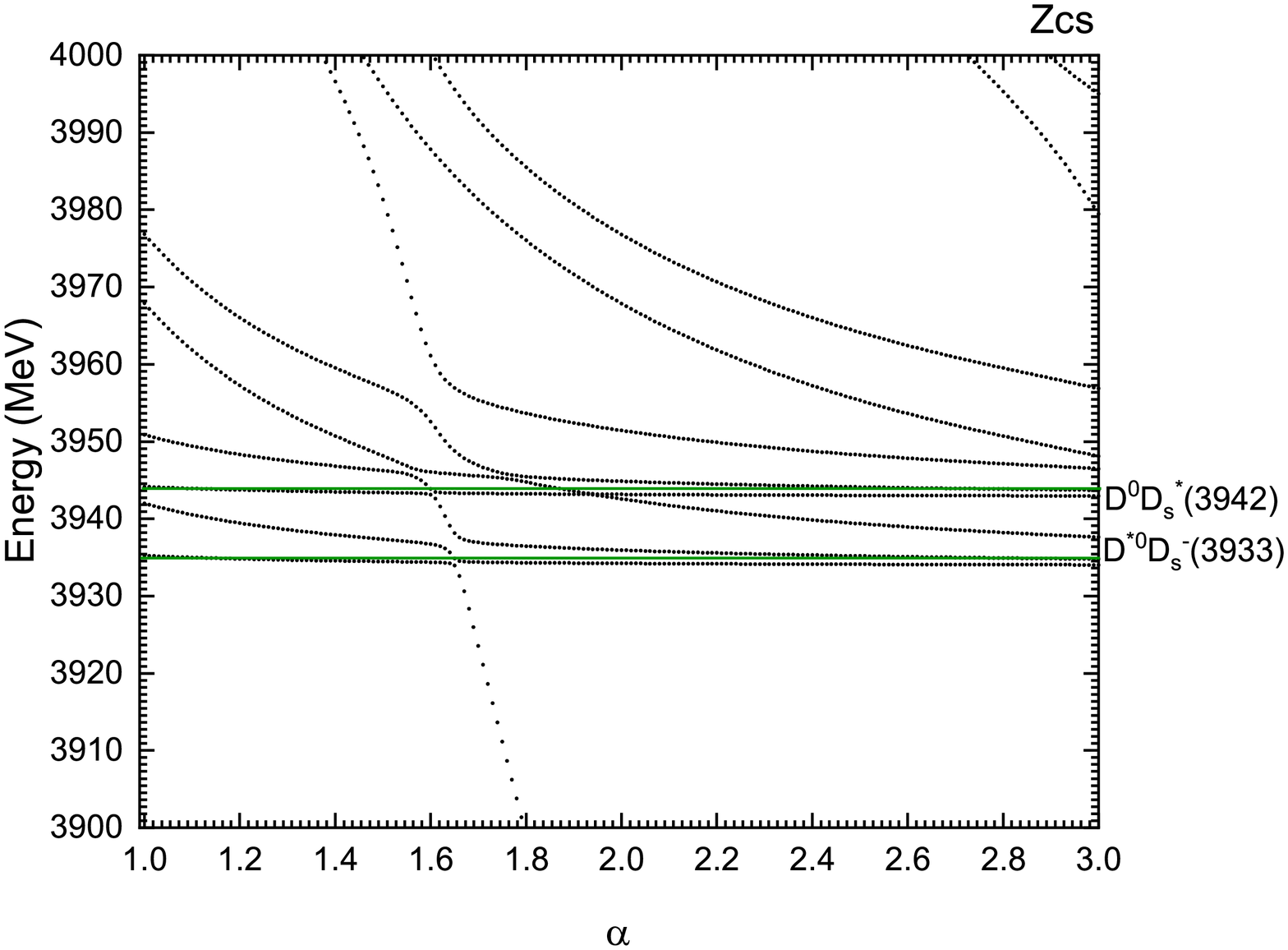}}
\caption{\label{Zcs2} The stabilization plots of the energies (3900 MeV $\sim$ 4000 MeV) of $c\bar{c}s\bar{u}$ states for $I(J^P)=\frac{1}{2}(1^+)$ with respect to the scaling factor $\alpha$.}
\end{figure}

\begin{figure}
\resizebox{0.50\textwidth}{!}{\includegraphics{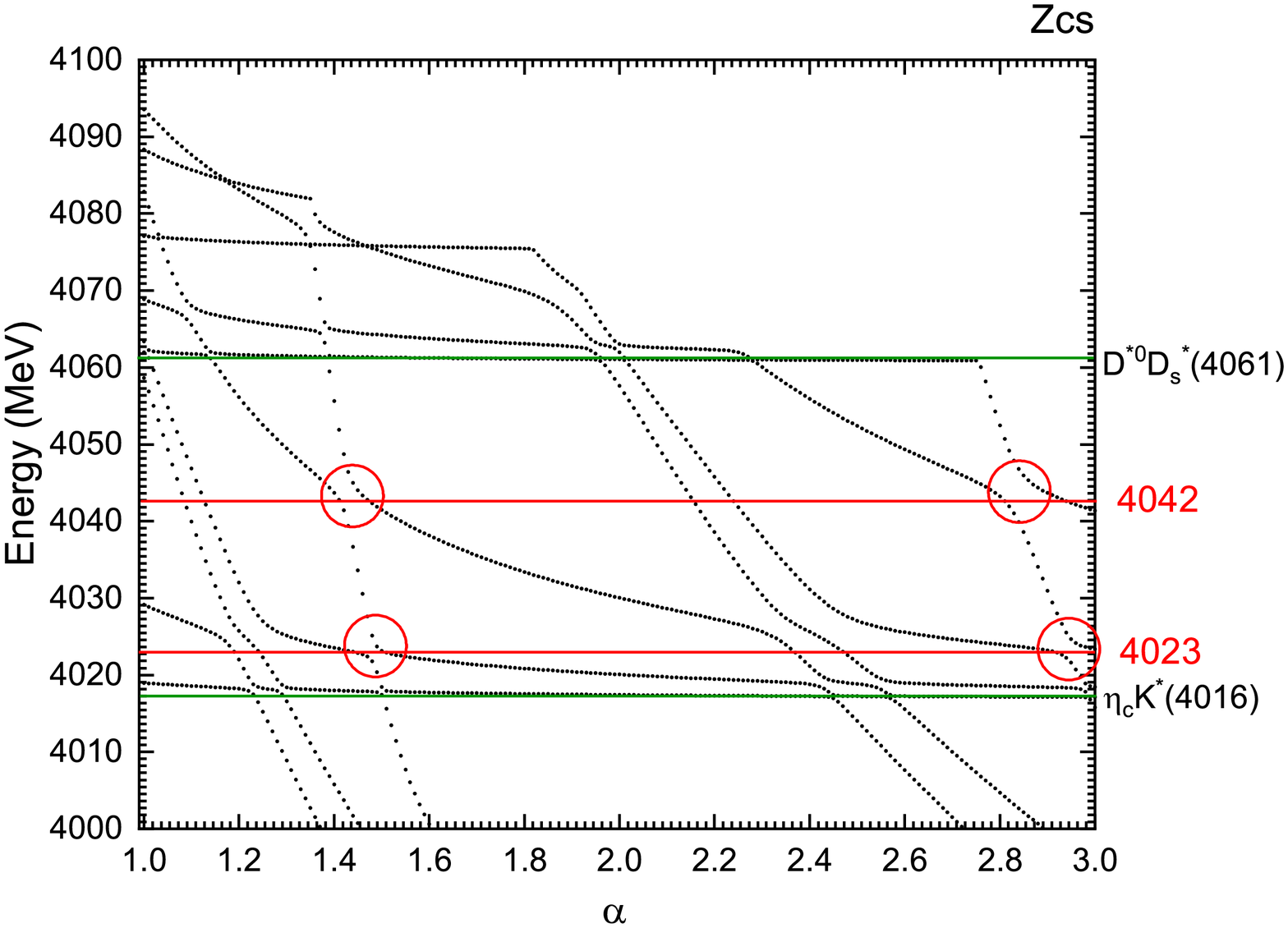}}
\caption{\label{Zcs3} The stabilization plots of the energies (4000 MeV $\sim$ 4100 MeV) of $c\bar{c}s\bar{u}$ states for $I(J^P)=\frac{1}{2}(1^+)$ with respect to the scaling factor $\alpha$.}
\end{figure}

\begin{figure}
\resizebox{0.50\textwidth}{!}{\includegraphics{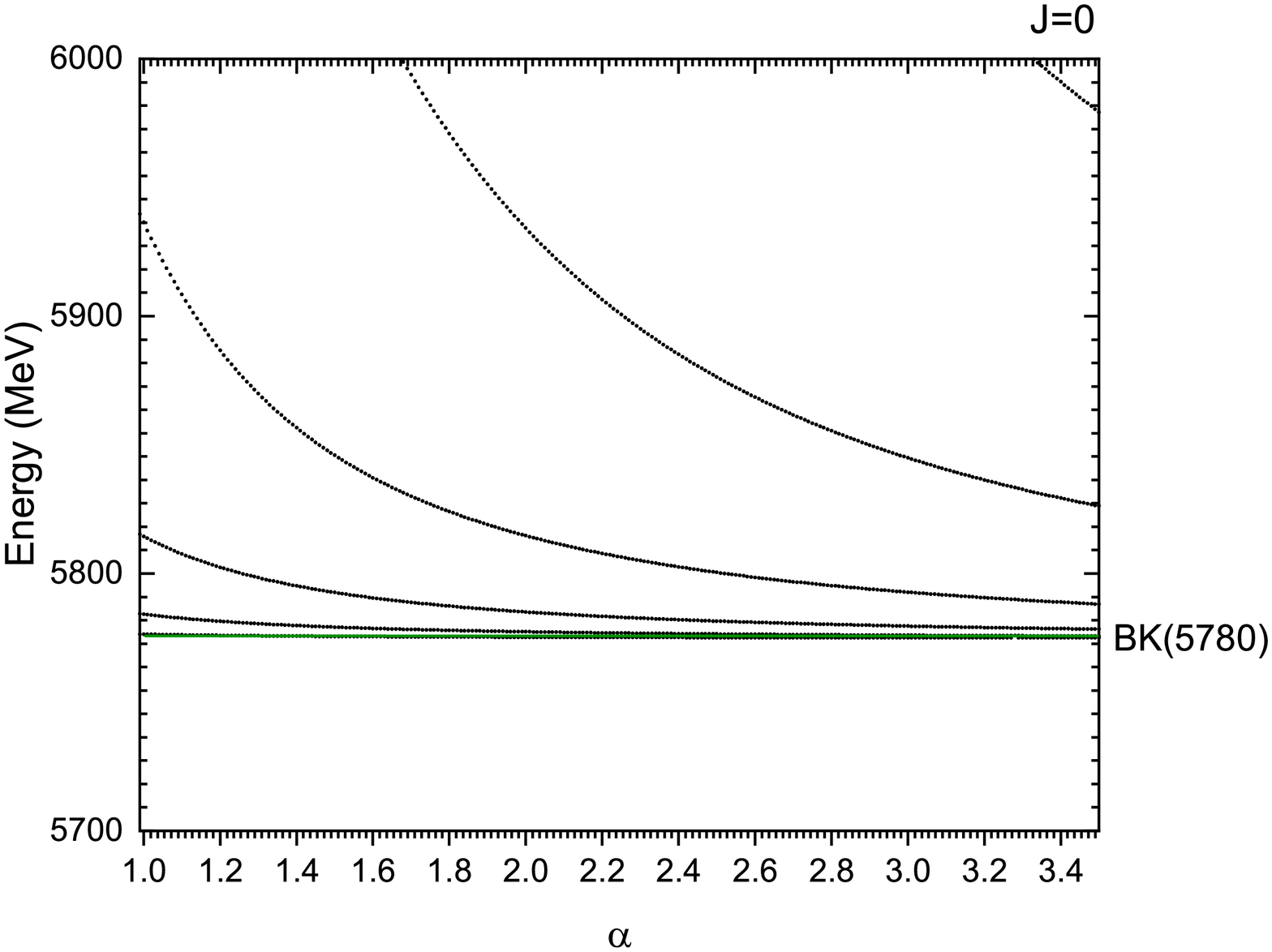}}
\caption{\label{bs01} The stabilization plots of the energies (5700 MeV $\sim$ 6000 MeV) of $b\bar{s}q\bar{q}$ states for $I(J^P)=0(0^+)$ with respect to the scaling factor $\alpha$.}
\end{figure}

\begin{figure}
\resizebox{0.50\textwidth}{!}{\includegraphics{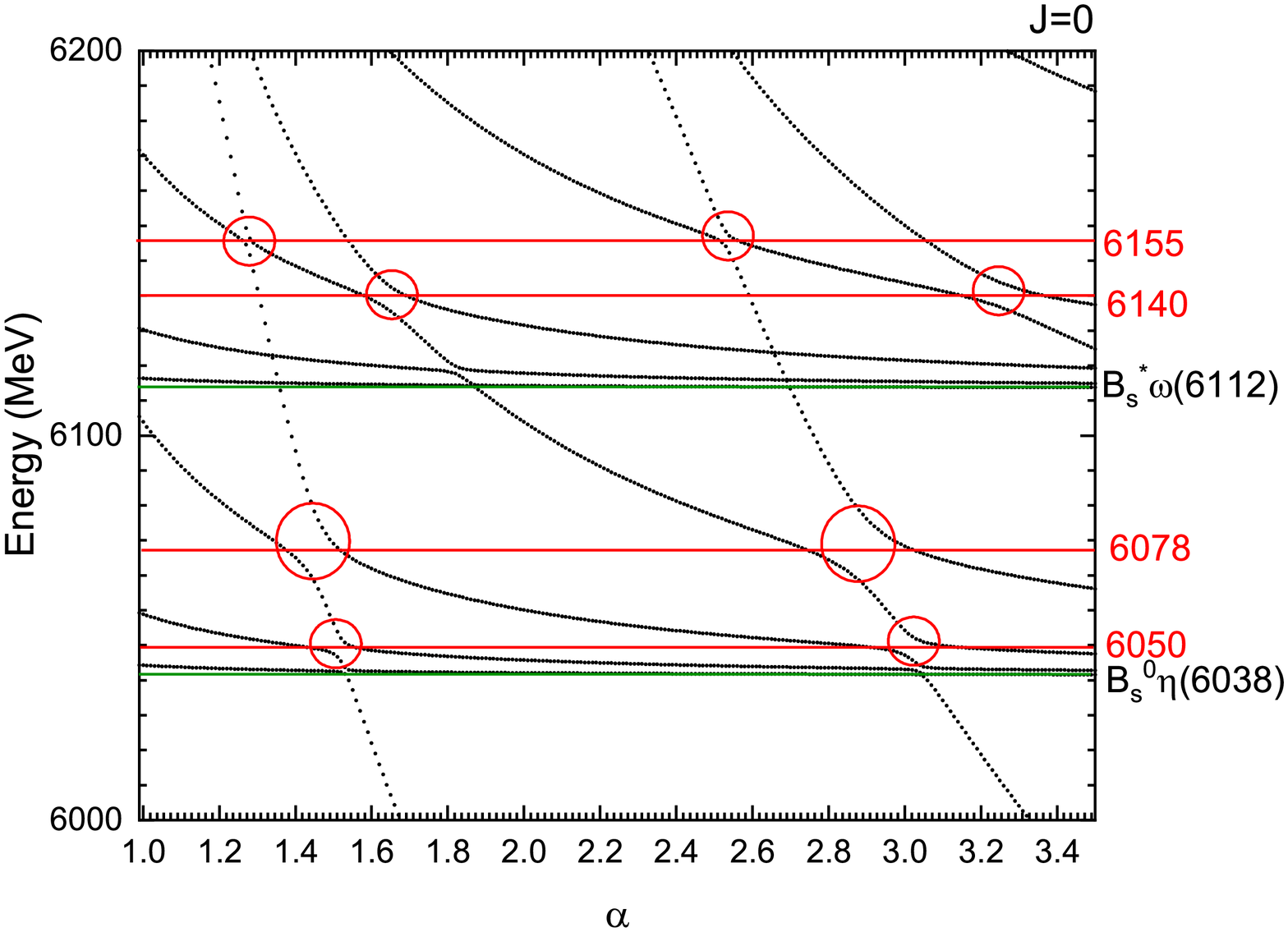}}
\caption{\label{bs02} The stabilization plots of the energies (6000 MeV $\sim$ 6200 MeV) of $b\bar{s}q\bar{q}$ states for $I(J^P)=0(0^+)$ with respect to the scaling factor $\alpha$.}
\end{figure}

\begin{figure}
\resizebox{0.50\textwidth}{!}{\includegraphics{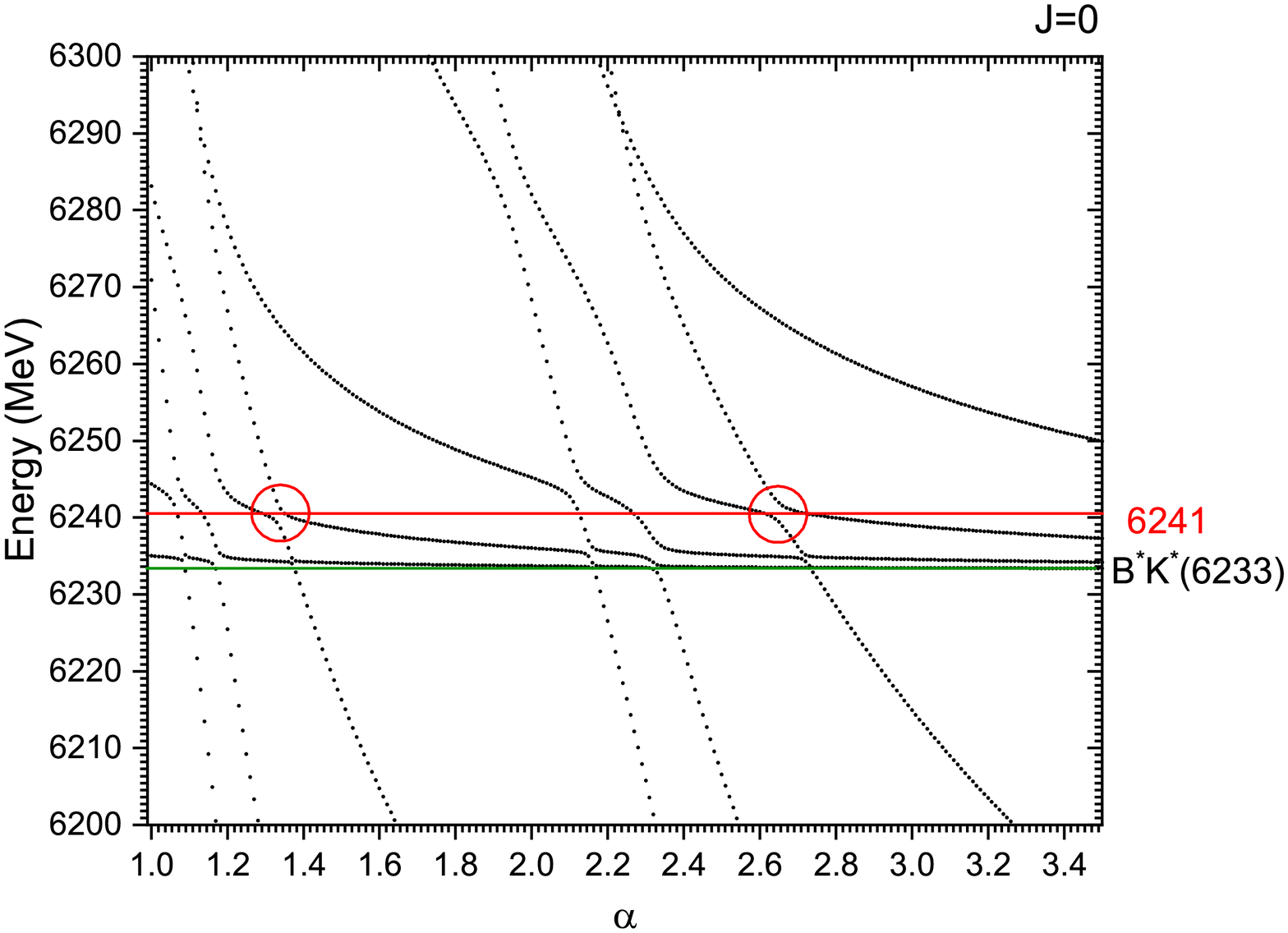}}
\caption{\label{bs03}The stabilization plots of the energies (6200 MeV $\sim$ 6300 MeV) of $b\bar{s}q\bar{q}$ states for $I(J^P)=0(0^+)$ with respect to the scaling factor $\alpha$.}
\end{figure}

\begin{figure}
\resizebox{0.50\textwidth}{!}{\includegraphics{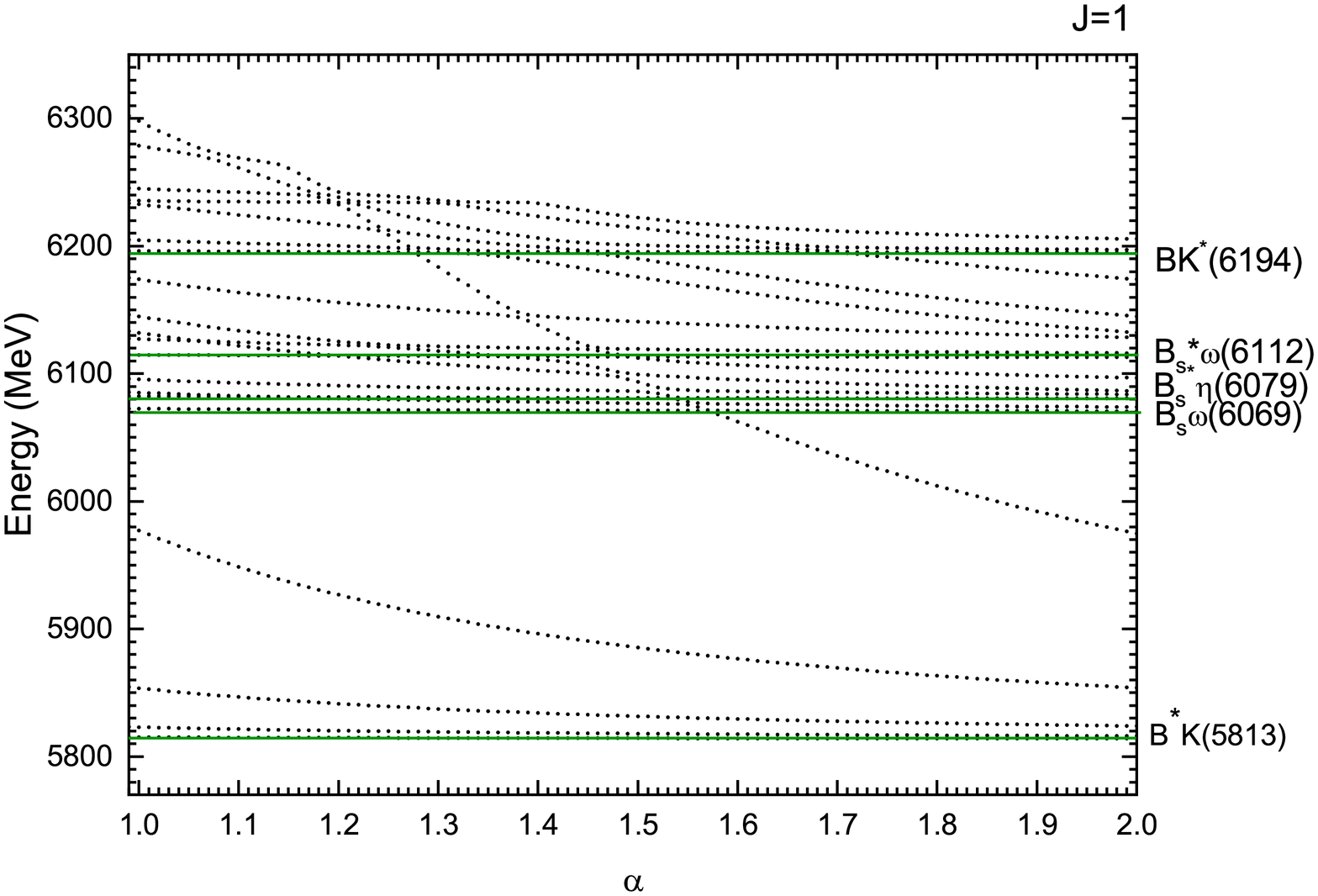}}
\caption{\label{bs1} The stabilization plots of the energies of $b\bar{s}q\bar{q}$ states for $I(J^P)=0(1^+)$ with respect to the scaling factor $\alpha$.}
\end{figure}

\begin{figure}
\resizebox{0.50\textwidth}{!}{\includegraphics{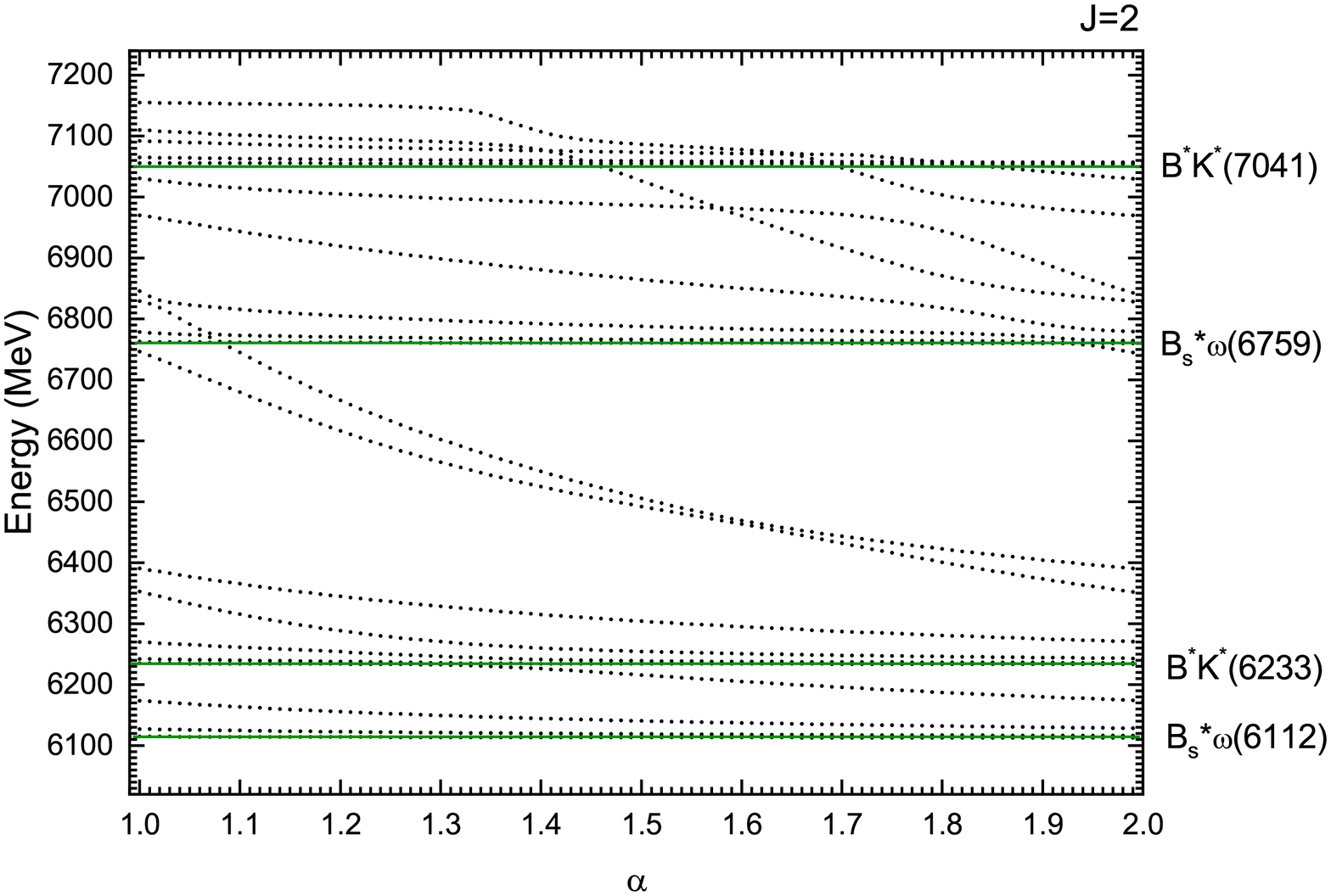}}
\caption{\label{bs2} The stabilization plots of the energies of $b\bar{s}q\bar{q}$ states for $I(J^P)=0(2^+)$ with respect to the scaling factor $\alpha$.}
\end{figure}

From the table, both for $Z_{cs}$ and $b\bar{s}q\bar{q}$ system, we can easily found that the low-lying energies in diquark-antidiquark
are all much larger than those in meson-meson structures. All of them are higher than the lowest theoretical thresholds. Besides, the effects of
the structures mixing seem to be tiny for the ground state energy. The coupling energies $E_{cc}$ are a little higher than the relevant
thresholds. So we cannot find the bound states of $c\bar{c}s\bar{u}$ and $b\bar{s}q\bar{q}$ tetraquark in the chiral quark model.

Because the colorful clusters cannot fall apart directly, there may exist resonances even with the higher energies. Using the
stabilization method (real scaling method), we try to find possible resonance for $c\bar{c}s\bar{u}$ and $b\bar{s}q\bar{q}$ system.
To realize the real scaling method here, we multiply the Gaussian size parameters $r_n$ in Eq. (\ref{gaussiansize}) by a factor
$\alpha$, $r_n \rightarrow \alpha r_n$ only for the meson-meson structure with color singlet-singlet configuration. Then we can locate the
resonances of $c\bar{c}s\bar{u}$ and $b\bar{s}q\bar{q}$ system with respect to the scaling factor $\alpha$, which takes the values from
1.0 to 3.5. With the variation of $\alpha$, the scattering states will level off to corresponding thresholds, but a resonance will appear as a avoid-crossing structure, which is illustrated in Fig. \ref{avoid-crossing} \cite{rs1}. The above line represents a scattering state,
and it will fall down to the threshold. The down line is the resonance state, which try to keep stable. The resonance state will interact with the scattering state, which can bring about a avoid-crossing point in Fig. \ref{avoid-crossing}. With the increasing of the scaling
factor $\alpha$, if we can observe repeated avoid-crossing point, it will be a resonance \cite{rs1}.

To make it clear for the reader, we illustrated the stabilization plots of the energies from 3600 MeV to 4100 MeV for $Z_{cs}$ states,
respectively in Fig. \ref{Zcs1}, Fig. \ref{Zcs2}, Fig. \ref{Zcs3}. In Fig. \ref{Zcs1}, we see the first green horizontal line,
which represents the lowest threshold $J/\psi K (1 \otimes 0 \rightarrow 1)$. In the energy region 3900 MeV to 4000 MeV (Fig. \ref{Zcs2}),
there are two thresholds $D^{*0}D_s^- (1 \otimes 0 \rightarrow 1)$ and $D^0D_s^* (0 \otimes 1 \rightarrow 1)$. In higher energy range
4000 MeV to 4100 MeV in Fig. \ref{Zcs3}, two thresholds $\eta_c K^* (0 \otimes 1 \rightarrow 1)$ and $D^{*0}D_s^* (1 \otimes 1 \rightarrow 1)$
appear. Meanwhile, in the figure, we can clearly see the repeated avoid-crossing points which are marked with red circles and the red horizontal
lines are on behalf of two genuine resonance states, with the energy 4023 MeV and 4042 MeV. The energies of the resonances are not far from
the experimental values of $Z_{cs}(3985)^-$ observed by BESIII and $Z_{cs}(4000)^+$ observed by LHCb.

For $b\bar{s}q\bar{q}$ system, we show the results with all possible quantum numbers $I(J^P)=0(0^+)$, $0(1^+)$ and $0(2^+)$ in
Figs. (\ref{bs01}-\ref{bs2}). Figs. (\ref{bs01})-(\ref{bs03}) represent the $b\bar{s}q\bar{q}$ system for $0(0^+)$. In the energy range
5700 MeV to 6000 MeV (Fig. \ref{bs01}), there is one threshold $BK$, and no resonance is found. Above 6000 MeV in Figs. \ref{bs02} and
\ref{bs03}, we find several resonances, such as 6050 MeV, 6078 MeV, 6140 MeV, 6155 MeV and 6241 MeV. From Fig. \ref{bs02}, we can found
that resonance states with energies 6140 MeV, 6155 MeV have the same resonant line. To identify which state is the real resonance state,
we calculate the proportions of total 12 channels for these two eigen-states. And we find that at the avoid-crossing point for state
with 6140 MeV, the color singlet channels $B_s^0\eta$ and $B_s^*\omega$ play a major role. But for state with 6155 MeV, the hidden-color
channels occupy an important role. So we abandon the state with energy 6140 MeV and the state with energy 6155 MeV is the real resonant state
in our calculation. For $0(1^+)$ and $0(2^+)$ states in Fig. \ref{bs1} and Fig. \ref{bs2}, we cannot find any resonance states in our work.

Besides, we calculated the decay widths of these resonance states using the formula taken from reference \cite{rs1},
\begin{align}
\Gamma=4|V(\alpha)|\frac{\sqrt{|S_r||S_c|}}{|S_c-S_r|},
\end{align}
where, $V(\alpha)$ is the difference between the two energies at the avoid-crossing point with the same value $\alpha$. $S_r$ and $S_c$
are the slopes of scattering line and resonance line, respectively. For each resonance, we get the decay width at the first and the second
avoid-crossing point, and we finally give the average decay width of these two values. The results are shown in Table \ref{decaywidth}.
For $c\bar{c}s\bar{u}(4042)$ state, the decay width of 13.7 MeV is very consistent with the experimental values of $Z_{cs}(3985)^-$, with decay width of 12.8 MeV. Besides, we can see that the mass and decay width of $b\bar{s}q\bar{q}~(6078)$ state are relatively close to the experimental values $M=6063$ MeV and $\Gamma=26$ MeV by LHCb Collobations \cite{Aaij:2020hcw}. Combining with the results of $b\bar{s}$ system, it is possible that the newly observed excited $B^0_s$ states are mixing states of $b\bar{s}$ and $b\bar{s}q\bar{q}~(q=u,d)$. The unquenched quark model should be invoked to study the highly excited mesons.

%\linespread{1.5}
\begin{table}[!t]
\begin{center}
\renewcommand\tabcolsep{6.0pt} % 调整表格列间的宽度
\caption{ \label{decaywidth} The decay widths of resonances of $c\bar{c}s\bar{u}$ and $b\bar{s}q\bar{q}$ system. (unit: MeV).}
\begin{tabular}{cccc}
\hline\hline\noalign{\smallskip}
Resonance State                         &$\Gamma$    &Resonance State  &$\Gamma$  \\
\hline
$c\bar{c}s\bar{u}$(4023) &3.1  &$c\bar{c}s\bar{u}$(4042) &13.7\\
$b\bar{s}q\bar{q}$(6050) &7.8  &$b\bar{s}q\bar{q}$(6078) &44.1 \\
$b\bar{s}q\bar{q}$(6155) &8.7  &$b\bar{s}q\bar{q}$(6241) &4.1 \\
\hline\hline
\end{tabular}
\end{center}
\end{table}

\section{Summary}
\label{epilogue}
Motivated by the recent experimental information from BESIII and LHCb Collaboration, we calculated the mass spectrum of the $Z_{cs}$
with $I(J^P)=\frac{1}{2}(1^+)$ and $B_s^0$ states with $I(J^P)=0(0^+), 0(1^+), 0(2^+)$ in the framework of the chiral quark model using
the Gaussian expansion method. Meson-meson and diquark antidiquark structures, and the coupling of them are considered.

For $Z_{cs}$ state with quark component $c\bar{c}s\bar{u}$, we found that the low-lying eigenvalues are all higher than the corresponding
thresholds in either structure, leaving no space for a bound state. But we found two resonances with mass 4023 MeV and 4042 MeV for
$c\bar{c}s\bar{u}$ system with the help of the real scaling method, and the decay width is 3.1 MeV and 13.7 MeV, respectively.
The state $c\bar{c}s\bar{u}(4042)$ has a consistent mass with the recent observed state $Z_{cs}(3985)^-$ and $Z_{cs}(4000)^+$, but
the decay width is close to the experimental value of $Z_{cs}(3985)^-$ and far narrower than the experimental value $Z_{cs}(4000)^+$.

To find the excited $B_s^0$ state observed by LHCb Collaboration, we give the mass spectrum both in 2-body $b\bar{s}$ system and 4-body
$b\bar{s}q\bar{q}(q = u~\rm{or}~d)$ system by considering the possible production of quark-antiquark pair in the vacuum. For quark-antiquark
system, the $2S$ and $1D$ states have masses close to the newly observed $B_s^0$, so the chiral quark model can accommodate these excited
$B_s^0$ states. For four-quark system, no bound state is found. However several resonances are emerged. They have energies, 6050 MeV,
6078 MeV, 6155 MeV, and 6241 MeV. The decay width are all relatively narrow, with 7.8 MeV, 44.1 MeV, 8.7 MeV and 4.1 MeV, respectively.
Comparing with the experimental data, we found that it is also possible to interpret the observed $B_s^0$ states as four-quark states. Therefore the better way to investigate the highly excited states is to invoke the unquenched quark model~\cite{Chen:2017mug}, which is our future work.

These possible resonant states should be tested in more precise experimental data in the future and we need more experimental studies
on the dominant decay channels of $Z_{cs}$ and $B_s$ to figure out their inner configurations.

\end{document}